\documentclass[twocolumn,showpacs,preprintnumbers,amsmath,amssymb]{revtex4}

\newcommand{\FileFig}[1]{#1}
\usepackage{graphicx}
\usepackage{dcolumn}
\usepackage{bm}

\begin{document}

\title{Quasi-particle tunneling at a constriction \\ in a fractional quantum Hall state}
\author{Stefano Roddaro}
\email{roddaro@sns.it}
\author{Vittorio Pellegrini}
\author{Fabio Beltram}
\affiliation{NEST-INFM and Scuola Normale Superiore, I-56126 Pisa, Italy}
\date{\today}


\begin{abstract}
Split-gate constrictions can be used to produce controllable scattering in a fractional quantum Hall state and constitute a very versatile model system for the investigation of non-Fermi physics in edge states. Controllable inter-edge tunneling can be achieved by tuning the constriction parameters and its out-of-equilibrium behavior can be explored as well. Here we review our results of tunneling non-linearities at a split-gate constriction in a wide range of temperatures and inter-edge coupling. The results are compared to available theoretical predictions of tunneling characteristics between Luttinger liquids. We show how partial agreement with these theoretical models is obtained in selected ranges of temperatures and inter-edge coupling, while striking deviations are obtained especially in the low-coupling, low-temperature regimes.
\end{abstract}

\maketitle


A two-dimensional electron system (2DES) in the fractional quantum Hall (FQH) regime can give rise to many ``exotic'' phenomena driven by electron-electron interactions \cite{QHE1}. Under the application of a strong magnetic field and for peculiar ratios of the charge ($n$) and flux quanta ($n_\phi=eB/h$) densities the 2DES undergoes transitions to insulating Hall phases. For ``magic'' values of the fractional filling factor $\nu=n_\phi/n$ the ground state excitations are expected to be fractionally charged and to obey fractional statistics \cite{Laughlin}. In both the integer and FQH effect, bulk states are characterized by an excitation gap while the only low-energy charged excitations propagate at the edge of the quantum Hall liquid thus creating a conducting, {\em chiral} one-dimensional (1D) channel. Wen predicted \cite{Wen1} that the edge of a FQH phase at $\nu=1/q$, where $q$ is an odd integer, should be completely equivalent to a Luttinger liquid (LL) with interaction parameter $g=\nu$. This peculiar model is known as {\em chiral} LL ($\chi$LL) and represents one of the simplest and most remarkable conceptual examples of a non-Fermi metal.

The FQH edge state is not the only electronic system that is expected to support a non-fermionic behavior. In recent years, the experimental realization of a LL has become a target of several research efforts that also explored different fabrication strategies of condensed matter systems. In particular cleaved-edge overgrowth (CEO) was used to produce clean and long quantum wires: temperature power laws as well as resonant transport were measured and compared with LL predictions \cite{Yacoby}. Finally carbon nanotubes are emerging as a promising model system for the verification of LL physics \cite{Ishii,GlattliCN}.

Edge FQH states are expected to lead to a particularly {\em robust} realization of a LL: edge channels at $\nu=1/q$ with $q$ odd integer, in fact, can only propagate in one direction and do not suffer from backscattering by random impurities. No localization occurs and the residual disorder only affects the electrons giving a mere phase shift. These issues stimulated much experimental effort that provided striking indications of non-Fermi physics at work for edge-states in the fractional QH regime \cite{Chang,Grayson,Yang03,Milliken,Goldman,Glattli,Roddaro1,Roddaro2,Roddaro3,Heiblum0,Heiblum1,Heiblum2}. 
Many discrepancies and inconsistencies with the $\chi$LL description however remain and the validation of the Luttinger liquid model is still not conclusive. These difficulties triggered a significant theoretical effort \cite{Raimondi,Papa04,Mandal01,Halperin,Koutouza,Saleur03, Mandal02,Chamon} aimed at improving the understanding of the impact of electron-electron interactions and edge structure on the non-Fermi liquid nature of FQH edge states. 

In this paper we present our tunneling experiments between fractional quantum Hall edge states. This approach constitutes one of the most direct ways to probe the properties of FQH edges. In order to experimentally open an inter-edge tunneling channel, a split-gate constriction (or quantum point contact) is fabricated on top of a FQH liquid to force the edge states to be in close proximity in a spatially-localized region. We exploit this to trigger the tunneling between otherwise non-interacting edge states. Transmission and reflection amplitudes of quasiparticles at the impurity are among the quantities that can be calculated with great accuracy. Our experiments can thus be used to test the validity of the approximations used in the theoretical models. In this paper we provide a detailed comparison with the predictions of the model proposed by Fendley, Ludwig and Saleur \cite{Fendley1} that provides a unified theoretical framework of non-equilibrium transport between $\chi$LLs applicable to different regimes of inter-edge coupling. 

The paper is organized as follows. In section I we present the experimental set-up and discuss the basic scheme for the measurement of tunneling at a split-gate constriction. The fundamental theoretical regimes of the weak-backscattering limit (WBL) and strong-backscattering limit (SBL) are also presented. Section II is devoted to the experimental results: we review our non-linear tunneling characteristics at different inter-edge coupling conditions and we emphasize the main differences between SBL and WBL. The evolution of the out-of-equilibrium tunneling characteristics as a function of the temperature is also presented. Experimental data are discussed in comparison with theoretical predictions. In section III we compare the experimental data with the results of a numerical approach described in \cite{Fendley1} that allows to calculate non-perturbatively the tunneling curves for all coupling regimes. 


\section{Inter-edge tunneling at a constriction}

\subsection{Measurement scheme} 

Quantum Hall systems in the presence of a tunable constriction provide a very useful implementation of a controllable scattering center in a LL. 

\begin{figure}
\includegraphics[width=0.48\textwidth]{\FileFig{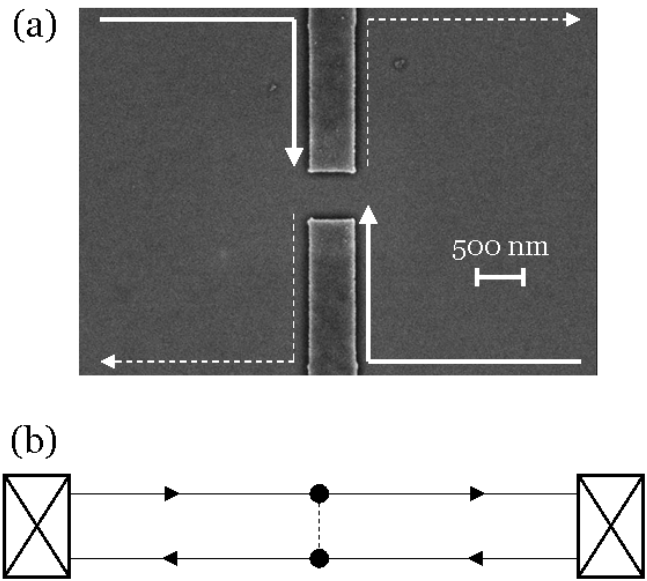}}
\caption{\label{fig:01} (a) Scanning electron picture of a representative split-gate structure fabricated by e-beam lithography. Geometrically the metallic fingers depicted in figure are $500\,{\rm nm}$ wide and they are separated by a $500\,{\rm nm}$ wide gap. The role of the constriction is to induce an inter-edge tunneling. Incoming edge states (continuous line) are scattered into outgoing edge states (dashed lines) in a controllable way. This system can be associated to an artificial, controllable scattering center (b) in a one-dimensional system. The width of the actual conducting channel far from the constrictions is much larger than depicted in figure ($\approx 80\,{\rm \mu m}$).}
\end{figure}

\begin{figure}
\includegraphics[width=0.48\textwidth]{\FileFig{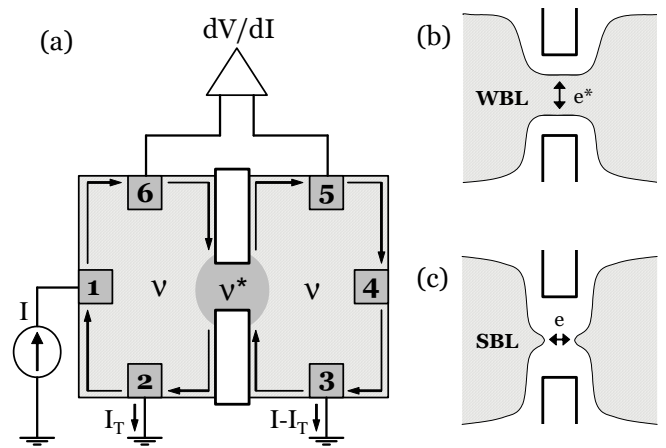}}
\caption{\label{fig:02} (a) A split-gate constriction in a FQH state constitutes a model system for the verification of the effects of a single impurity on a chiral edge state. It has been suggested that at filling factors $\nu=1/q$, where $q$ is an odd integer, edge states are equivalent to a single-branch Luttinger Liquid \cite{Wen1}. Two opposite limits can be identified in the inter-edge tunneling evolution. (b) Weak backscattering limit (WBL): edge states are nearly not-interacting and the tunneling is expected to be driven by the transfer of fractional quasiparticles of charge $e^*$ between two different portions of the edge of the quantum Hall phase. (c) Strong backscattering limit (SBL): edge states are nearly completely reflected by the constriction. A completely different tunneling geometry is realized and tunneling is associated to scattering of {\em electrons} between two disconnected quantum Hall regions.}
\end{figure}

In our experiments (see Fig.~\ref{fig:01}), inter-edge tunneling is induced by a local constriction imposed on the 2DES: edge states are forced to flow in a small localized region and back-scattering can be induced in a controllable way. Semiconductor structures studied in the experiments here presented were fabricated from high-mobility AlGaAs/GaAs single heterojuctions or quantum wells. We analyzed different samples with sheet density in the range $0.7-1.0\times 10^{11}\,{\rm cm^{-2}}$ and electron mobility always exceeding $10^{6}\,{\rm cm^2/Vs}$. We used different 2DES with a depth going from $100$ to $300\,{\rm nm}$. In our devices the constrictions were defined electrostatically by means of split-gate electrodes. Figure~\ref{fig:01} reports a scanning electron micrograph of a representative device fabricated by e-beam lithography, evaporation and lift off of composite layers of Al and Au. In Fig.~\ref{fig:01}a, the rectangular regions correspond to the metallic layer. When gated to a negative voltage bias value $V_g$ these metallic electrodes locally deplete the 2DES and produce the constriction. Typical dimensions of our QPCs ($300-500\,{\rm nm}$ wide fingers separated by a gap of $500-600\,{\rm nm}$) allow the formation of a {\em wide} passage where edge states can still flow without experiencing a significant back-scattering. By further lowering the gate bias $V_g$ it is possible to shrink the constriction down to pinch-off. In this scheme, $V_g$ is the parameter used to tune the inter-edge coupling strength. At the same time $V_g$ also controls the electron density in the region shaped by the split-gate potential. This results into a reduced local density $n^*$ and a reduced filling factor $\nu^*$ at the constriction region. 

Figure~\ref{fig:02}(a) reports the measurement scheme. By means of out-of-equilibrium four-wire differential resistance measurements it is possible to bias the inter-edge tunneling junction and to measure the tunneling characteristics. The biasing is achieved by injecting a current $I$ (a small ac component plus a dc offset) at contact $\emph 1$. As a consequence, the potential difference between edge states arriving to the QPC from the top-left and from the bottom-right is 

\begin{equation}
V_T=\rho_{xy}I.
\label{eq:00}
\end{equation}

This potential corresponds to the tunneling bias. The current impinging the QPC is partially transmitted and reflected and is collected at contacts $\emph 2$ and $\emph 3$. We will refer to the reflected tunnel current as $I_T$ while the transmitted signal corresponds to $I-I_T$. The redistribution of the currents at the edge gives, after thermal equilibration, different voltage drops. It is easy to demonstrate that a differential measurement between contacts $5$ and $6$ yields

\begin{equation}
\delta V_{56}=\rho_{xy}\delta I_T.
\label{eq:01}
\end{equation}

As a consequence, the differential resistance measured following the scheme of Fig.~\ref{fig:02} directly relates to the differential reflection coefficient ($dI_T/dI$) and to the differential backscattering conductance

\begin{equation}
\frac{dV}{dI}=\frac{dV_{56}}{dI}=\rho_{xy}\frac{dI_T}{dI}=\rho_{xy}^2\frac{dI_T}{dV_T}.
\label{eq:02}
\end{equation}

Measurements can also be performed in a simple four-wire setup, i.e. without grounding contact $\emph 2$. These results can still be interpreted following equation (\ref{eq:02}) as long as the system is in the WBL, where the effect of contact $\emph 2$ is not relevant. For higher tunneling rates the exact mapping of the measured differential resistance into tunneling curves is still possible but becomes less direct \cite{Note01}.

\subsection{Tunneling regimes and non-linearity} 

Theoretically, the tunneling between edge-states displays a complex evolution as a function of the inter-edge coupling strength that can be captured with non-perturbative approaches \cite{Fendley1}. We address this in the last section. It is also possible, however, to tackle the problem considering the two limit cases. In the WBL, edge states are nearly not-interacting. In this case a small interaction leads to quasiparticle (with charge $e^*=e/q$) tunneling and to very peculiar I-V characteristics. The strong inter-edge correlations causes an enhanced tunneling (i.e. back-scattering) at zero bias and low-temperature (in the $T=0$ limit the {\em so-called} overlap catastrophe \cite{KaneFisher} is obtained). As consequence a peculiar differential conductance ($dI_T/dV_T$) peak is expected at {\em zero bias} \cite{Wen1}. Peak amplitude and width are predicted to scale as $T^{2\nu-2}$ and $T$, respectively. When the tunneling current increases the system enters a complex mixed regime till the SBL is finally reached. In this latter limit edge states are strongly coupled and at low-temperature most of the current impinging on the constriction is back-scattered to the counter-flowing edge state and reflected back to the source. In order to analyze this case it is useful to start from the complete reflection limit and study perturbatively the forward tunneling. Following the available theory \cite{KaneFisher} this {\em forward-scattering} is associated to the transfer of electrons between two disconnected QH regions separated by the pinched-off constriction. For simple fractions such as $\nu=1/q$ where $q$ is an odd integer, the differential conductance for the forward tunneling vanishes at zero bias with a peculiar power law $I\propto V^{2q-1}$. In this regime, tunneling non-linearities can be described as a direct consequence of the modulation of the density of states for the electrons in the LL ($\propto(E-E_F)^{q-1}$). This regime was explored in pioneering experiments in CEO structures where tunneling between a metallic region and an edge state was induced \cite{Chang,Grayson}. 

As stated above, a complete non-perturbative treatment spanning throughly the tunneling between the WBL and SBL was reported in \cite{Fendley1}. In section III we will give more details about the theoretical evolution of these tunneling curves and their comparison with experimental data.


\section{Experimental results}

\subsection{Wide constrictions ($\nu^*\approx\nu$)}

\begin{figure}
\includegraphics[width=0.48\textwidth]{\FileFig{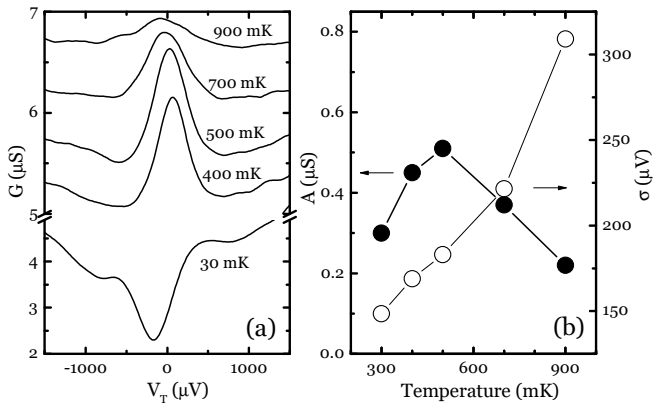}}
\caption{\label{fig:03} Panel (a): evolution of backscattering curves as a function of temperature. At high temperture a  peak is observed. As temperature is lowered at first the peak grows, but for $T<300\,{\rm mk}$ the curve reverts to an unexpected {\em minimum}. Panel (b): fundamental parameters (peak amplitude $A$, and peak width $\sigma$ as obtained from a gaussian fit) of curves reported on the top part of panel (a).}
\end{figure}

Figure~\ref{fig:03}(a) reports a set of tunneling conductance curves at a fixed gate voltage (just above the 2D-1D threshold) and different temperatures as a function of the tunneling bias $V_T$. Traces are shifted vertically for clarity. As the temperature is increased above $300\,{\rm mK}$ a strong {\em zero bias} back-scattering peak is reported. This behavior is in qualitative agreement with the theory developed in Ref.\cite{Wen1} that applies to the WBL limit and constitutes an nice evidence in support of inter-edge quasiparticle tunneling at the constriction. In order to have a quantitative analysis we also report, in panel (b), the most important peak parameters (as obtained from a gaussian fit), i.e. the peak width and amplitude. We find out that the peak width \cite{Note05} is $\approx40\%$ wider that expected from \cite{Wen1}. 

An unexpected behavior emerges when the temperature is lowered to values below approximately $300-400\,{\rm mK}$. The back-scattering peak develops into a {\em minimum} (see bottom curve in panel (a) of Fig.~\ref{fig:03}). This behavior, which seems in agreement with data reported in other experiments \cite{Heiblum2}, cannot be described in terms of available theories in either the SBL or WBL which indicate that backscattering should be strongly enhanced at low temperatures. These discrepancies add to some unexpected results found in experimental studies of CEO systems \cite{Chang} and pose the question about the universality of the inter-edge tunneling in the FQH regime. These findings have stimulated additional theoretical analysis that suggested possible origins for these discrepancies. The new research directions are focussing either on the effect of the inter-edge coulomb interaction \cite{Raimondi}, on the residual interaction between the composite fermions \cite{Mandal01}, on the internal structure of the real edge state \cite{Mandal02}, and on the different tunneling geometries induced by the actual scattering center \cite{Papa04}.


\begin{figure}
\includegraphics[width=0.45\textwidth]{\FileFig{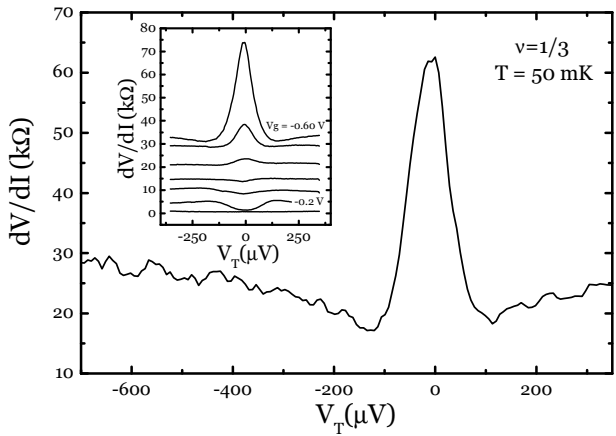}}
\caption{\label{fig:04} Low temperature evolution of the tunneling curves. For a relatively pinched constriction a backscattering peak is obtained also at low temperature. The inset displays the evolution of the conductance curves as a function of the split-gate bias (i.e. of the inter edge coupling strength).}
\end{figure}

\subsection{Approaching pinch-off ($\nu^*<\nu$)}

An interesting evolution can also be observed as the constriction parameters are changed and one moves far away from the WBL. Figure~\ref{fig:04} reports a tunneling peak obtained at $50\,{\rm mK}$ for a higher inter-edge interaction obtained by negatively biasing the split-gate that induces the constriction. The inset reports a representative evolution of the tunneling curves as a function of the split-gate bias. When the edge states are weakly interacting (and $\nu^*\approx\nu$) a {\em reduced} tunneling is observed at zero bias (curve corresponding to $V_g=-0.2\,{\rm V}$ in the inset of Fig.~\ref{fig:04}), in agreement with data reported in Fig.~\ref{fig:03}. This lineshape reverts again to a clear backscattering peak at higher interaction strengths (i.e. at lower, more negative values of $V_g$). This evolution was observed in different samples. The presence of a reduced filling factor $\nu^*$ inside the constriction region could be relevant to the understanding of this behavior \cite{Roddaro3}. A more detailed analysis of this backscattering regime will be discussed in the next section.


\section{Comparison with theory, beyond the WBL and SBL}

In the previous sections we discussed the SBL and WBL. They give a nice and rather intuitive description of the inter-edge tunneling in proximity to the fixed points $G=0$ and $G=\nu e^2/h$, where $G$ is the two-wire conductance of the system. In our experiment, however, most of the results belong to regimes that are typically in between the WBL and SBL. It is thus important to consider models that allow to calculate the tunneling curves at arbitrary coupling strengths. In the first part of this section we therefore introduce the approach proposed by Fendley, Ludwig and Saleur \cite{Fendley1} and following this model we provide calculations of the tunneling conductance that can be directly compared to experimental data. This model proposes an exact solution to the problem of transport through LL in presence of a single impurity. Its application to FQH states is based on the theoretical framework proposed by Wen \cite{Wen1} and Kane and Fisher \cite{KaneFisher} that is able to describe some but not all the observed experimental features. For this reason the understanding of the degree of applicability of the model by Fendley et al. to the description of tunneling in the FQH state can be considered still an open issue. To address this question, however, a detailed comparison between experimental tunneling curves and the corresponding predictions of the theory is required. In the second part of the section we discuss this comparison with the experimental data shown in the previous sections and provide the evolution of the calculated tunneling behavior that applies to the description of results close to the SBL. 

Fendley et al. exploited the fact that the LL model with an impurity is {\em integrable} to calculate transport properties of such a system non-perturbatively in the impurity strength. In this approach, the authors defined a quasiparticle basis in which the scattering problem can be solved in a Landauer-B\"uttiker-like spirit. Using thermodynamic Bethe ansatz techniques (refer to \cite{Fendley1} for a detailed description) the non-linear conductance $G(V,T,T_B)$ can be calculated explicitly for $\nu<1$ and any bare value of the impurity strength. The conductance is a universal function of the relative {\em ratios} of $V$ (tunneling voltage), $T$ (temperature) and $T_B$ (the only free parameter of the theory: this temperature models the impurity strength). $G$ also depends on the value of $g$, but we will restrict here to the case of interest for us, i.e. $g=1/3$ (also curves at $g=1/5$ will be presented in Fig.~\ref{fig:05}). In view of the experimental data, it is important to consider that $G$ corresponds in the model to the {\em transmission} conductance at the constriction QPC: differential backscattering conductance (the quantity addressed in all of our measurements) can be simply obtained by $e^2/3h - G$.

\begin{figure}
\includegraphics[width=0.4\textwidth]{\FileFig{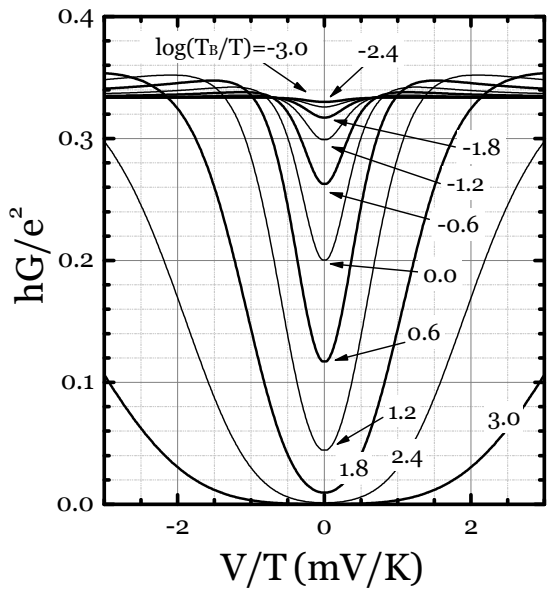}}
\caption{\label{fig:fen01} Theoretical differential conductance ($G$) curves at different interaction strength $T_B/T$. In the limit $T_B/T\rightarrow0$ the curves tend to Wen's prediction \cite{Wen1}. For higher values of $T_B/T$ the SBL is achieved and curves collapse on the $T=0$ curve $G_{T=0}(V/T_B)$.}
\end{figure}

Figure~\ref{fig:fen01} reports the sequence of universal curves $G$ vs. $V/T$ at various relative tunneling strength $T_B/T$. The plot reports tunneling curves ranging from the WBL (corresponding to curves at $G\approx e^2/3h$, where a perfect overlap with Wen's predictions \cite{Wen1} is obtained) to very high tunneling strengths. In order to really follow the SBL of the theory, it is convenient to plot the curves as a function of $V/T_B$. As $T_B/T$ diverges all the curves collapse on a $T=0$ limit curve $G_{T=0}(V/T_B)$ (Fig.~1 of \cite{Fendley1} reports the curve). In this limit the theory correctly captures the power laws $G\propto V^4$ and $G\propto T^4$.

In the next paragraph we shall compare these predictions with the temperature evolution of the tunneling back-scattering peaks reported in the previous sections.

\subsection{Experimental lineshapes}

\begin{figure}
\includegraphics[width=0.45\textwidth]{\FileFig{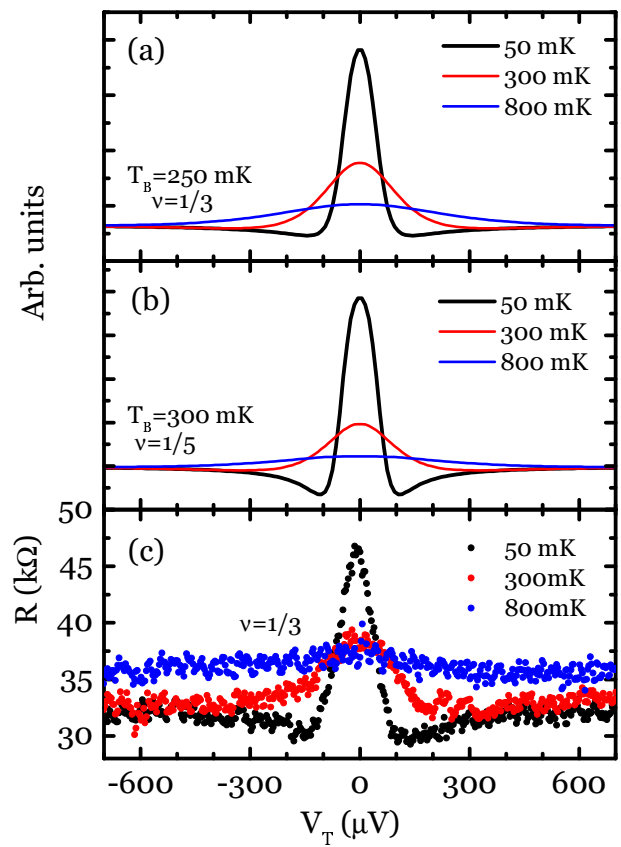}}
\caption{\label{fig:05} Comparison between theoretical lineshapes and experimental data. Panel (a) and (b): curves obtained following \cite{Fendley1}. $T_B$ is the only real free parameter of the theory. We plotted both the conductance for an edge state at $\nu=1/3$ and $\nu=1/5$ (equivalent to a LL with $g=\nu$ \cite{Wen1}). Panel (c): experimental curves obtained with a bulk filling factor $\nu=1/3$ in a relatively pinched constriction. The lineshape and its temperature evolution present a remarkable similarity with theoretical prediction. Data have been obtained using a current modulation of $20\,{\rm pA}$, corresponding to about $1.5\,{\rm \mu V}$ in terms of tunneling voltage ($V_T$).}
\end{figure}

Figure~\ref{fig:05}(c) reports the temperature evolution for a tunneling peak obtained for a rather strongly pinched constriction (in a condition similar to the one of data in Fig.~\ref{fig:04}). Panel (a) and (b) report a comparative plots of theoretical curves obtained following \cite{Fendley1}. Experimental lineshape and its temperature evolution display remarkable similarities with theoretical predictions. In particular, the two lateral minima of the data at $50\,{\rm mK}$ represent a non-trivial out-of-equilibrium feature that has a nice correspondence with theoretical predictions.

It is very important to note how the actual backscattering background (about $32\,{k\Omega}$ \cite{Note06}) and the scale of the peak's amplitudes are not easy to compare with theory. Panel (a) reports the theoretical differential back-scattering between two edge states at $\nu=1/3$: the curves should display a {\em zero} background back-scattering with amplitude ranging between $0$ and $\rho_{xy}\approx 77.5\,{\rm k\Omega}$. By assuming that the filling factor in the constriction region is reduced, however, the observed back-scattering background could be explained in terms of {\em $\nu$-matching} at the boundary of the constriction. This leads to a finite value of the longitudinal resistance at large $V_T$'s. The value of 32$k\Omega$ observed in the data of Fig.~\ref{fig:05}(c) corresponds to a reduced filling factor $\nu^*$ = 1/5. For this reason we also reported in panel (b) the theoretical lineshape for $\nu=1/5$. In addition it should be noted that the peak value at zero-bias was found to change even in the same sample and depending on the cool-down procedure. This suggests that the absolute value of the zero-bias peak is affected by random impurities that perturb the zero-bias tunneling strength within the constriction region.  

\subsection{On the range of validity of SBL predictions}

\begin{figure}
\includegraphics[width=0.48\textwidth]{\FileFig{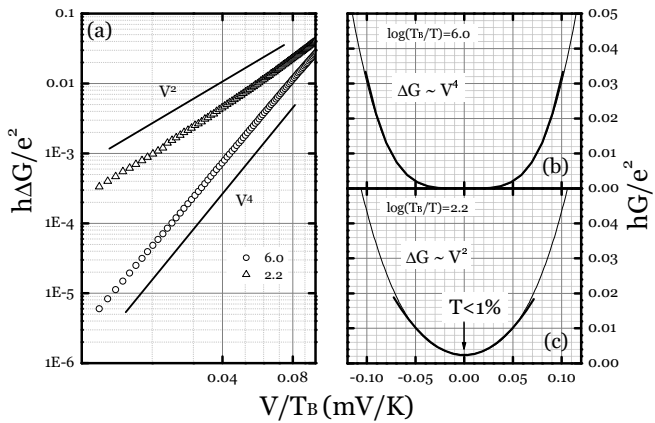}}
\caption{\label{fig:fen02} Curves that intuitively belong to the SBL can still display striking departures to the predicted $\Delta G \propto V^4$ power law. Panel (a): log-log plot of $\Delta G(V)=G(V)-G(0)$ for a comparison between the tunneling curves at $\log(T_B/T)=2.2$ (panel (b)) and $\log(T_B/T)=6.0$ (panel (c)). For $\log(T_B/T)=2.2$ (minimum transmission $<1\%$) the conductance curve is still, to a very good approximation, simply {\em parabolic} and strikingly deviates from the approximate SBL prediction. At higher tunneling strengths the SBL power law $V^4$ is finally, and correctly, obtained.}
\end{figure}

The very signatures of the SBL in the inter-edge tunneling are the predictions $G\propto V^4$ and $G\propto T^4$ (for $\nu=1/3$). This is tipically interpreted as a consequence of a peculiar density of states at the Fermi energy $\propto (E-E_F)^2$. This very intuitive behavior, however, is a {\em limit} behavior and can only hold in close proximity of the $G=0$ fixed point. The exact point at which this limit description breaks down is not easy to determine. 

It is interesting to stress that, from inspection of curves reported in Fig.~\ref{fig:fen01}, it turns out that the leading power law for $G$ remains $V^2$ down to very low forward-tunneling rates. Curves with peak reflection of $99\%$ are still, to a very good approximation, simply {\em parabolic} \cite{Note02}. Figure~\ref{fig:fen02} reports the conductance curves corresponding to $\log(T_B/T)=2.2$ and $\log(T_B/T)=6.0$. The log-log plot in panel (a) highlights the different power laws for $\Delta G(V)=G(V)-G(0)$. This analysis could be relevant for the interpretation of experimental data in the SBL such as the ones reported in \cite{GlattliPE}. In that article a significant deviation from the $V^4$ dependence was shown. The SBL behavior only emerges very close to the $G=0$ limit. This gives some useful indications about the effective limits for the observation of the SBL power laws: even in the simplest case of a single impurity in a perfect LL (which already does not necessarily provide an exact description of the constriction problem for an edge state at $\nu=1/3$), striking deviations from $\Delta G \propto V^4$ can be easily obtained.


\section{Conclusions}

Tunneling at a split-gate constriction in a fractional quantum Hall state constitute a unique tool to probe the nature of the correlated electron edge-state. Remarkable out-of-equilibrium non-linear behavior emerges that supports the notion of a non-Fermi nature of the edge state in the FQH regime. An comprehensive theoretical description of these features, however, is far from giving a complete understanding of the physics associated to the backscattering events at a constriction. In this paper, we reviewed our recent experiments on quasiparticle tunneling at a constriction and we provided a critical analysis of these data in terms of the available theories. The picture that emerges is intriguing and shows many open fundamental issues still to be understood. 
\section{Acknowledgements}

We thank G. Biasiol and L. Sorba of the TASC laboratory in Trieste for the growth of semiconductor samples. We are also grateful to  R. D'Agosta, M. Grayson, J.K. Jain, A.H. MacDonald, E. Papa, R. Raimondi, B. Trauzettel, G. Vignale for discussions and suggestions. This work was supported in part by the Italian Ministry of University and Research under FIRB RBNE01FSWY and by the European Community's Human Potential Programme under contract HPRN-CT-2002-00291 (COLLECT).

\end{document}